\documentstyle[twocolumn,prl,aps,epsf]{revtex}
\begin{document}
\def \beq{\begin{equation}}
\def \eeq{\end{equation}}
\def \beqarr{\begin{eqnarray}}
\def \eeqarr{\end{eqnarray}}

\twocolumn[\hsize\textwidth\columnwidth\hsize\csname @twocolumnfalse\endcsname
\draft

\title{
Quantum Theory 
of Pomeranchuck Transition in 
Two Dimensional Fermi Liquids via High Dimensional Bosonization
}

\author{Kun Yang}
\address{
NHMFL and Department of Physics,
Florida State University, Tallahassee, Florida 32306, USA
}

\date{\today}

\maketitle

\begin{abstract}

We use high dimensional bosonization to derive 
an effective field theory that describes the Pomeranchuck 
transition in two-dimensional Fermi liquids.
The bosonization approach explicitly retains all low-energy
degrees of freedom of the system.
The resultant theory
has dynamical exponent $z=2$ at tree level and 
upper critical dimension $d_c=2$,
thus in 2D the system is at the upper critical dimension.  
These results differ from those of
an earlier study based on integrating out fermions.

\end{abstract}
]

{\em Introduction and Motivation} ---
Quantum phase transitions\cite{sachdev}
are of tremendous current interest to the physics community in general and
condensed matter physicists in particular. 
Significant progress has been made in theoretical studies of
quantum phase transitions in
various spin and boson models\cite{sachdev}; examples include various
magnetic transitions in insulating spin systems, and superfluid-insulator
transitions in bosonic systems. Historically however, the first examples of
quantum phase transitions studied theoretically are ferromagnetic
and antiferromagnetic transitions in {\em itinerant} electron systems, or
metals\cite{hertz}. Ironically, quantum phase transitions in itinerant
electron systems turn out to be among the most difficult to study theoretically,
and our current understanding of these transitions is much more limited
compared to those of spin and boson systems. The origin of the difficulty is
not hard to see. Just like classical phase transitions, theoretical studies of
quantum phase transitions are based on (quantum versions) of Ginsburg-Landau
type of free energy functionals, and their analyzes based on Wilsonian
renormalization group (RG). The Ginsburg-Landau free energy is a functional of
the order parameter, which are {\em bosonic} degrees of freedom; fluctuations
of their long-wave length, low-frequency/energy
components determine the critical
properties of the transitions. In itinerant electron systems
however, the fundamental low-energy degrees of freedom are the gapless
electronic states
near the Fermi surface; these {\em fermonic} modes at {\em finite}
wavevectors are difficult to incorporate within the Ginsburg-Landau-Wilson 
(GLW) paradigm. In the approach pioneered by Hertz\cite{hertz} and extended by
Millis\cite{millis} and others, one decouples the electron-electron
interaction using Hubbard-Stratanovish transformation in an appropriate
channel, integrates out the
fermionic degrees of freedom,
and arrives at a GLW-like free energy functional
that involves the Hubbard-Stratanovish auxiliary field only,
which are bosonic degrees
of freedom and interpreted as the order parameter.
It has been realized recently, however, that the procedure of integrating
out gapless fermions leads to singularities in the expansion of the
resultant bosonic free energy functional in terms of the order parameter, or
its gradients\cite{bk}. Such singularities may invalidate the standard
classification and analyzes (based on power-counting)
of various terms of the GLW-like theory obtained this way,
and profoundly affect the critical behavior of the
transitions\cite{bk}. The presence of such singularities is a
consequence of integrating out gapless degrees of freedom, 
and also reflects the
fact that the GLW-like theory obtained this way only provides
an incomplete description of the low-energy physics of the system.
At present there is no consensus on the appropriate
approach to study quantum phase transitions in itinerant electron systems
in general.

Recently, we
studied the ferromagnetic transition in one-dimensional (1D) itinerant
electron systems, using a different approach\cite{yang}.
The idea was quite simple: In 1D one can use the very powerful machinery of
bosonization, which allows one to obtain a {\em bosonic} description of the
{\em fermionic} system, {\em without} losing any fermionic degrees of freedom;
{\em i.e.}, one can represent the fermionic degrees of freedom using bosonic
degrees of freedom faithfully, via bosonization.
Extensive previous studies have established that the paramagnetic phase in 1D
is described by a {\em free} boson theory, known as the Luttinger liquid
theory in this context;
this is a consequence of the fact that all
possible bosonic
interactions are {\em irrelevant} at the Luttinger liquid fixed point,
and scale to zero in the low energy limit.
We have shown\cite{yang}, on the other hand,
at the Gaussian critical point of the ferromagnetic
transition, interactions
are {\em relevant}, and must be included in the analyzes of the critical
behavior, as well as the ordered phase.
We have performed such analyzes using RG, and found the behavior
of the system at the
ferromagnetic critical point
is quite different from that of the Luttinger liquid,
due to
the presence of interaction; this ``non-Luttinger liquid" behavior
is in close correspondence with the experimentally
observed non-Fermi liquid behavior near magnetic transitions in higher
dimensions. It is worth
emphasizing that the theory we developed using bosonization does {\em not}
suffer
from the singularities encountered in the Hertz-Millis approach in high
dimensional systems discussed
above, since no low-energy fermionic degrees of freedom are lost\cite{kim}.

The bosonization machinery
has been generalized to higher dimensional systems\cite{haldane,hm,fradkin,kopietz}. 
In general bosonization is not as powerful in high D as in 1D. This is
because to bosonize the fermions,
one needs to divide the Fermi surface (which is
a continuous manifold in high D in contrast to discrete points in 1D)
into many small patches that are almost flat; the scattering processes
that bring the electron from one patch to another show up as non-linear
operators within bosonization, rendering them difficult to treat. What the
bosonization approach does treat well are the forward scattering processes
which leave the electrons in the {\em same} patch that they start from, as they
are represented by quadratic terms of the boson field; these 
are precisely the Fermi liquid interactions. For this reason bosonization gives
a very good description of the Fermi liquid phase\cite{haldane,hm,fradkin,kopietz}.

In the present work we use high D bosonization to study the Pomeranchuck 
transition in 2D spinless
Fermi liquids, a quantum phase transition at which 
the Fermi surface develops anisotropy spontaneously, which is of considerable
current interest\cite{vadim,other,barci}. The bosonization 
approach is particularly useful here because the Pomeranchuck
transition is driven by 
{\em Fermi liquid interactions}. 
The strategy here is similar to Ref. \onlinecite{yang}.
Our central result is the GLW theory for the
transition, Eq. (\ref{action}). From the action (\ref{action}) we can readily 
read out that the tree-level 
dynamical exponent $z=2$ at the transition,
and upper critical dimension $d_c=4-z=2$. Thus in
2D the system is at the upper critical dimension. This differs from the result
an earlier study\cite{vadim} based on Hertz-Millis type of approach, which 
found $z=3$ and $d_c=1$. We will discuss the origin of the difference of these
two approaches.

{\em Bosonized Description of Fermi Liquids} ---
We start by reviewing the bosonized description of Fermi liquids, which also 
sets the notation for the rest of the paper.
As the first step we divide the Fermi surface into many small
patches of linear size $\Lambda\ll k_F$, 
so that within each patch the Fermi surface is approximately flat,
and for electronic states sufficiently close to the Fermi surface 
(within a cutoff distance $\lambda\ll k_F$), the 
energy is approximately linear in momentum. For each patch we introduce a 
patch density operator
\beq
\rho({\bf S}, {\bf q})=\sum_{\bf k}\theta({\bf S}, {\bf k})
c^\dagger_{{\bf k}-{\bf q}}c_{\bf k},
\eeq
where ${\bf S}$ is the patch label, and $\theta({\bf S}, {\bf k})$ is 1 when
${\bf k}$ is inside the box of size $\Lambda^{D-1}\lambda$ enclosing patch
${\bf S}$ ($D$ is the dimensionality of the system), 
and zero otherwise; it ensures we are summing over states near
patch ${\bf S}$ only. They satisfy the bosonic commutation relation
\beq
[\rho({\bf S}, {\bf q}), \rho({\bf T}, {\bf q}')]=\Omega
\delta_{{\bf S},{\bf T}}\delta_{{\bf q},-{\bf q}'}({\bf q}\cdot\hat{n}_{\bf S}),
\label{comm}
\eeq
with
$\Omega=\Lambda^{D-1}({L/2\pi})^D$,
and $L$ is the linear size of the system and $\hat{n}_{\bf S}$ is the unit 
vector pointing in the outward normal direction of patch ${\bf S}$.
In the case of 1D the Fermi surface patch reduces to a discrete Fermi point, and
Eq. (\ref{comm}) reduces to the familiar 1D commutator of density operators.

The foundation of bosonization lies on the fact that both the kinetic energy 
and the Fermi liquid (or forward scattering)
interaction terms of the electron Hamiltonian can be expressed in terms of
$\rho({\bf S}, {\bf q})$. Within the approximation that
the kinetic energy is linear in momentum within each patch:
$\epsilon_{\bf S}({\bf k}) \approx v_F({\bf S})\hat{n}_{\bf S}
\cdot[{\bf k}-{\bf k}_F({\bf S})]$, where $v_F({\bf S})$ is the Fermi velocity of patch ${\bf S}$,
the kinetic energy is quadratic in $\rho({\bf S}, {\bf q})$:
\beq
T=\sum_{{\bf S}, {\bf q}}[v_F({\bf S})/2\Omega]\rho({\bf S}, {\bf q})
\rho({\bf S}, -{\bf q}).
\eeq 
Non-linearity in the spectrum leads to higher order terms in 
$\rho({\bf S}, {\bf q})$; they are irrelevant for the description of the 
Fermi liquid fixed point, 
but are important for the description of the transition
and will be discussed later on. The Fermi liquid (or forward-scattering)
interaction takes the form 
\beq
V={1\over 2}\sum_{{\bf S}, {\bf T}, {\bf q}}V_{{\bf S}, {\bf T}}({\bf q})
\rho({\bf S}, {\bf q})\rho({\bf T}, -{\bf q}),
\label{V}
\eeq
where $V_{{\bf S}, {\bf T}}({\bf q})$ is the forward scattering matrix element
between states in patches ${\bf S}$ and ${\bf T}$; at the Fermi liquid fixed
point its dependence on ${\bf q}$ is irrelevant and we can take the ${\bf q}=0$
value in the long-wave length limit (here we do not consider possible
singularity due to
long-range interaction). The Fermi liquid Hamiltonian 
$H=T+V$,
which is expressed exclusively in terms of the patch density operators
$\rho({\bf S}, {\bf q})$, 
also has an equivalent Lagrangian description:
\beq
L=L_0\{\rho({\bf S}, {\bf q})\}-H\{\rho({\bf S}, {\bf q})\},
\label{lag}
\eeq
where the dynamical term\cite{fradkin,barci}
\beq
L_0={i\over 2\Omega}\sum_{{\bf S}, {\bf q}}[\partial_t\rho({\bf S}, {\bf q})]
\rho({\bf S}, -{\bf q})/({\bf q}\cdot\hat{n}_{\bf S})
\label{l0}
\eeq
properly enforces the commutation relation (\ref{comm}).

{\em Two-dimensional Isotropic Fermi Liquid} --- We now turn our discussion to
the case D=2, and assume isotropy so that the Fermi surface is a circle. In
this case the patch ${\bf S}$ can be labeled by an angular variable $\theta$,
and in the limit $\Lambda\ll k_F$, $\theta$ may be treated as a continuous
variable ranging between 0 and $2\pi$. In this case it is natural to perform
a Fourier transformation with respect to $\theta$:
\beqarr
\rho_m({\bf q})&=&{1\over 2\pi}\int{d\theta}e^{-im\theta}\rho(\theta, {\bf q});\\
\rho(\theta, {\bf q})&=&\sum_{m=-\pi k_F/\Lambda}^{\pi k_F/\Lambda}
e^{im\theta}\rho_m({\bf q}).
\eeqarr
The cutoff in the range of 
$m$ is due to the small but finite patch size $\Lambda$.
We can now express Eq. (\ref{l0}) in terms of $\rho_m({\bf q})$:
\beqarr
&L_0&=(ik_F/2\Lambda\Omega)\nonumber\\
&\times&\sum_{{\bf q},m,n}[\partial_t\rho_m({\bf q})]
\rho_{-n}(-{\bf q})\int{d\theta}e^{i(m-n)\theta}/[{\bf q}\cdot\hat{n}(\theta)]
\nonumber\\
&=&{i\pi k_F\over\Lambda \Omega}
\sum_{{\bf q},m}{[\partial_t\rho_m({\bf q})]\over|{\bf q}|}
\sum_{{\rm odd}\hskip 0.1cm l} (-1)^{{l-1\over 2}}e^{il\theta_{\bf q}}
\rho_{-m-l}(-{\bf q}),
\label{L0}
\eeqarr
where $\theta_{\bf q}$ is the angle of 2D vector ${\bf q}$.
It is very important to notice that in (\ref{L0}) $\rho$'s with even $m$ are 
coupled to $\rho$'s with odd $n$ {\em only}, and vice versa. 
This is because the integral
over $\theta$ vanishes when $m-n$ is even, so in the last line 
of (\ref{L0}) the sum over $l$ is for odd integers.

On the other hand it is easy to show that both $T$ and $V$ are diagonal in $m$,
and the Hamiltonian takes the form
\beq
H={\pi v_F k_F\over \Lambda\Omega}
\sum_{m,{\bf q}}(1+F_m)\rho_m({\bf q})\rho_{-m}(-{\bf q}),
\label{ham}
\eeq
where the dimensionless Landau parameter
\beq
F_m=({L\over 2\pi})^2{k_F\over v_F}\int_0^{2\pi}e^{im\theta}V(\theta),
\eeq
in which $V(\theta)$ is defined through 
$V_{{\bf S}, {\bf T}}({\bf q})=V(\theta_{\bf S}-\theta_{\bf T}, {\bf q})$,
and we take ${\bf q}=0$ here in the long wave-length description of Fermi 
liquid; the ${\bf q}$ dependence of $V$ will become important and discussed
later on. The stability of the isotropic Fermi liquid phase requires 
$F_m > -1$\cite{fradkin}.
Thus combining (\ref{L0}) and (\ref{ham}) we obtain a quadratic 
theory in terms of complex bosonic variables $\rho_m({\bf q})$, that describes
an isotropic 2D Fermi liquid. Notice that $L_0$ involves one time derivative
of $\rho_m({\bf q})$; this indicates $\rho_m({\bf q})$ are constrained degrees
of freedom (or their conjugate momenta are not independent variables). To bring
$L$ closer to the more familiar 1D bosonized description of Luttinger liquids,
we can integrate out $\rho_m({\bf q})$ with odd $m$, and obtain the Lagrangian
with even $m$ degrees of freedom:
\beqarr
&&L_{even}=L_{even}^0-H_{even},
\label{leven}
\\
&&L_{even}^0={A_e\pi k_F\over v_F\Lambda \Omega}\times\nonumber\\
&&\sum_{\bf q}{1\over |{\bf q}|^2} 
\sum_{m,n \hskip 0.05cm {\rm even}}
e^{i(m-n)(\theta_{\bf q}+{\pi\over 2})}
\partial_t\rho_{-m}(-{\bf q})\partial_t\rho_n({\bf q}),
\label{l0even}
\eeqarr
where 
$A_e=\sum_{l \hskip 0.05cm {\rm odd}} {1\over 4(1+F_l)}$
and $H_{even}$ takes the same form of (\ref{ham}) but with sum over even $m$.
Despite the fact that we have integrated out modes with odd $m$,
$L_{even}$ is completely equivalent to $L$ because $L^0_{even}$ contains two
time derivatives and thus the conjugate momenta of $\rho_m$ are independent 
degrees of freedom, which actually correspond to $\rho_m$ with odd $m$.
Alternatively we may choose to integrate out  modes with even $m$ to obtain a
dual description $L_{odd}$; this is analogous to the case in 1D where we may 
write down the Luttinger liquid Lagrangians in terms of either density or 
current fields, which are symmetric and antisymmetric combinations of left 
and right moving fields respectively.
Clearly $L$ and $L_{even}$ are scale invariant field theories with dynamical 
exponent $z=1$, because there is a factor $1/|{\bf q}|$ associated with each 
time derivative (or $\omega$ in frequency space) while $\theta_{\bf q}$ is 
invariant under scale transformations.

{\em The Pomeranchuck Instability and Ginsburg-Landau-Wilson Theory for 
the Transition} ---
Obviously the isotropic Fermi liquid phase becomes unstable when $F_m$ reaches
and goes below $-1$ for any $m$; this is the Pomeranchuck instability.
Let us assume this occurs (without losing generality) 
in an even channel $m_0\ne 0$. In this case we need to retain the
next leading quadratic term in $H$ for this channel, of the form 
\beq
V'=a\sum_{\bf q}|{\bf q}|^2\rho_{m_0}({\bf q})\rho_{-m_0}(-{\bf q})+\cdots;
\label{V'}
\eeq
it originates from the ${\bf q}$ dependence of $V_{{\bf S}, {\bf T}}({\bf q})$
in Eq. (\ref{V}), neglected at the isotropic Fermi liquid fixed point 
due to its irrelevance there. Here we assume $a > 0$.
To maintain stability, 
we also need to keep non-quadratic terms\cite{barci} of the form
\beqarr
&&T'=b\sum_{m,n,{\bf q}, {\bf q}'}\rho_m({\bf q})\rho_n({\bf q}')\rho_{-m-n}
(-{\bf q}-{\bf q}')\nonumber\\
&&+c\sum
\rho_m({\bf q})\rho_n({\bf q}')\rho_l({\bf q}'')
\rho_{-m-n-l} (-{\bf q}-{\bf q}'-{\bf q}'')\nonumber\\
&&+\cdots,
\label{T'}
\eeqarr
whose origin is the nonlinearity of electron dispersion near the Fermi surface
($b\propto\epsilon''(k_F)$ and $c\propto\epsilon'''(k_F)$),
again neglected at the isotropic Fermi liquid fixed point
due to its irrelevance. To proceed we focus for the moment on the $m_0$ 
channel where the instability occurs, and neglect its coupling to other
channels. From Eqs. (\ref{l0even}, \ref{V'}, \ref{T'})
we obtain the following GLW effective field theory with Euclidean
action
\beqarr
&S&=\int{d^D{\bf q}d\omega}\{({\omega^2\over |{\bf q}|^2}+|{\bf q}|^2+r)|\phi({\bf q},\omega)|^2
\nonumber\\
&+&(-1)^{m_0}{\omega^2\over |{\bf q}|^2}Re[e^{2im_0\theta_{\bf q}}\phi({\bf q},\omega)
\phi(-{\bf q},-\omega)]\}\nonumber\\
&+&u\int{d^D{\bf x}d\tau}|\phi({\bf x},\tau)|^4 + \cdots, 
\label{action}
\eeqarr
where $\phi\propto \rho_{m_0}$ is a complex bosonic field that plays the role
of order parameter of the theory, the ``mass"-like parameter
$r\propto 1+F_{m_0}$, and $u\propto c$ (also assumed positive); 
proper rescaling of the field as well
as space-time coordinates have been performed to ensure the form 
taken by the quadratic terms in Eq. (\ref{action}). This highly non-local 
action is somewhat similar to the one studied by Sachdev and Senthil 
(see Eq. 4.11 of Ref. \onlinecite{ss}) in a different context; the crucial 
difference here are the new
quadratic terms in the second line of Eq. (\ref{action}), 
which encode the information about the symmetry properties of the order
parameter as well as the Fermi surface dynamics. Clearly at the Gaussian
critical point $r=0$, we have dynamical exponent $z=2$ from the $\omega$ and
${\bf q}$ dependence of the quadratic part of the action. The tree-level 
flow equation for the interaction is simply given by its dimensionality:
\beq
{du\over d\log s}=(4-z-D)u,
\eeq
where $s$ the scaling parameter for spatial coordinates; we thus find the
upper critical dimension of the theory to be $d_c=4-z=2$. Higher order 
couplings are irrelevant at $D=2$.
These are the central results of this paper.

We now show that couplings to the other (non-critical) channels do not
change the form of the action (\ref{action}), or the associated critical
behavior, for the following reasons. 
(i) For the other channels the action contains a non-zero
mass-like term $(1+F_m)\rho_m({\bf q})\rho_{-m}(-{\bf q})$. 
Under the $z=2$ scaling the ``mass" $1+F_m$
is a relevant operator that grows as $s^2$ under scaling\cite{note}. 
Thus in the long-wave length limit all other channels become ``infinitely 
massive" and thus drop out of the low-energy effective action eventually.
(ii) Before this limit is reached, one needs to integrate over modes in these
channels with wave-vectors at the cutoff as RG proceeds; 
this generates various new couplings in 
the $m_0$ channel, in addition to renormalizing the existing ones.
Due to the presence of the mass-like terms, this is a well-behaved procedure, 
and due to the nature of the coupling in 
Eqs. (\ref{leven},\ref{l0even}) between the $m_0$ and other channels, 
the only possible
singular new couplings have positive powers of $\omega^2/|{\bf q}|^2$
attached to them, with no other singular dependence on $\omega$ or ${\bf q}$.
Under $z=2$ scaling, $\omega$ scales to zero faster than $|{\bf q}|$, and as a
consequence all such couplings are irrelevant. 

Normally
one would conclude that the critical behavior of the transition is
mean-field like with logarithmic corrections, based on
the fact that we are at the upper critical dimension. While this may very well 
be the case here, it should be noted that the quadratic terms of the action
(\ref{action}) are quite unconventional, which may lead to unconventional
flow to the interaction $u$ beyond tree level. We plan to study this as well as
electronic properties of the system near the transition in future work. 

In an earlier work\cite{vadim}, Oganesyan, Fradkin and Kivelson
obtained a bosonic theory for the transition from isotropic to nematic
Fermi liquids
by integrating out fermions via Hubbard-Stratanovich decoupling,
similar to the Hertz-Millis theory. This transition
corresponds to case $m_0=2$ in our
theory. These authors found $z=3$ and $d_c=1$. The
origin of $z=3$ lies in Landau damping, which is a 
consequence of integrating out
gapless fermions. In our
approach there is no analog of Landau damping. Technically this is because 
we do not integrate out any gapless modes. More fundamentally however, it is 
clear that the transition is driven by a bosonic {\em eigen} 
mode whose energy goes through zero; by definition eigen modes are not damped. 
In the presence of interactions
with other degrees
of freedom of the systems, the wave function of this unstable mode needs to 
be determined self-consistently, 
and RG is a systematic procedure 
that allows the long wavelength mode to adjust itself 
to remain an eigen mode as one approaches the low-energy limit. 
On the other hand if one integrates out the gapless fermions up-front, 
no adjustment is allowed and the physically relevant modes appear 
damped. It thus appears to be an artifact of this procedure.

This work was supported by NSF grant DMR-0225698.


\begin{references}
\bibitem{note} These terms are marginal under the $z=1$ scaling appropriate
for the Fermi liquid fixed point; since $z$ changes at the Pomeranchuck
transition, their scaling behavior changes as well.
\end{references}
\end{document}